\documentclass{article}

\PassOptionsToPackage{numbers, compress}{natbib}
\usepackage[final]{neurips_2023}

\usepackage[utf8]{inputenc} % allow utf-8 input
\usepackage[T1]{fontenc}    % use 8-bit T1 fonts
\usepackage{hyperref}       % hyperlinks
\usepackage{url}            % simple URL typesetting
\usepackage{booktabs}       % professional-quality tables
\usepackage{amsfonts}       % blackboard math symbols
\usepackage{nicefrac}       % compact symbols for 1/2, etc.
\usepackage{microtype}      % microtypography
\usepackage{xcolor}         % colors

\usepackage{subcaption} %我为子图引入的
\usepackage{graphicx}
\usepackage{amsmath}
\usepackage{multirow}

\title{WavMark: Watermarking for Audio Generation}

\author{%
	Guangyu Chen$^\dag$, Yu Wu$^\ddag$, Shujie Liu$^\ddag$, Tao Liu$^\dag$, Xiaoyong Du$^\dag$, Furu Wei$^\ddag$
    \\  Microsoft Research Asia$^\ddag$
  \\ Renmin University of China$^\dag$
}

\begin{document}

\maketitle
\begin{abstract}

Recent breakthroughs in zero-shot voice synthesis have enabled imitating a speaker's voice using just a few seconds of recording while maintaining a high level of realism. 
Alongside its potential benefits, 
this powerful technology introduces notable risks, including voice fraud and speaker impersonation.
Unlike the conventional approach of solely relying on passive methods for detecting synthetic data, watermarking presents a proactive and robust defence mechanism against these looming risks. 
This paper introduces an innovative audio watermarking framework that encodes up to 32 bits of watermark within a mere 1-second audio snippet. 
The watermark is imperceptible to human senses and exhibits strong resilience against various attacks. 
It can serve as an effective identifier for synthesized voices and holds potential for broader applications in audio copyright protection.
Moreover, this framework boasts high flexibility, allowing for the combination of multiple watermark segments to achieve heightened robustness and expanded capacity.
Utilizing 10 to 20-second audio as the host, 
our approach demonstrates an average Bit Error Rate (BER) of 0.48\% across ten common attacks, 
a remarkable reduction of over 2800\% in BER compared to the state-of-the-art watermarking tool. 
See \url{https://aka.ms/wavmark} for demos of our work.

% while maintaining a Signal-to-Noise Ratio (SNR) surpassing 36 dB.
% Utilizing a 20-second audio clip as the host, 
% our approach demonstrates an average Bit Error Rate (BER) of 0.37\% across ten different attacks, 
% while maintaining a Signal-to-Noise Ratio (SNR) surpassing 36 dB.
% See \url{https://aka.ms/wavmark} for demos of our work.

\end{abstract}

% \begin{figure}[th]
% 	\includegraphics[width=14cm]{figures/0_overview_v7}
% 	\centering
% 	\caption{
% Illustration of the encoding and decoding process of our framework.
% In the encoding phase, we iteratively embed a 1-second watermark segment into the host audio, 
% ensuring the entire time domain protection.
% %Even if the watermarked audio is clipped, we can still decoding, as long as there is a complete watermark segment in the clipped audio.
% %我们的模型支持从剪切的水印音频中进行提取。
% %Our model supports decoding from clipped watermarked audio.
% Since the watermarked audio may be clipped, 
% in the decoding process, we need to determine the  watermark position.
% To address this issue, we made our model to be shift-invariant, allowing for a maximum misalignment of 0.1 seconds in the decoding position.
% This characteristic enables us to efficiently apply a Brute Force Detection (BFD) strategy for decoding.
% Specifically, the encoded message is composed of pattern bits and payload bits. When decoding, we utilize a decoding window with a sliding step of 0.05 seconds to attempt decoding continuously.
% Then, the correct watermark results are identified by checking with the predefined pattern bits.
% 	} 
% 	\label{fig:overview}
% \end{figure}

\begin{figure}[thb]
	\includegraphics[width=14cm]{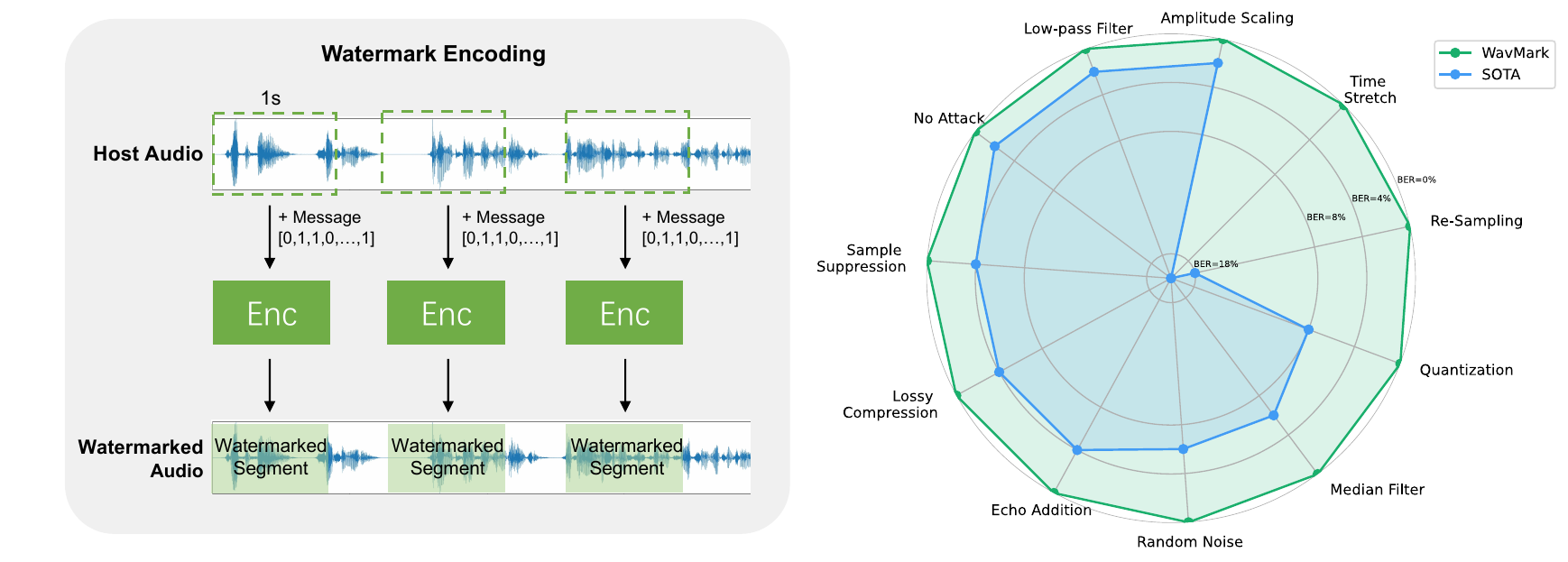}
	\centering
	\caption{
\textbf{Left:} The watermark encoding process of our framework. 
We iteratively add the same watermark into 1-second segments of the host audio to ensure full-time region protection.
Even if the watermarked audio is clipped,
decoding is possible using any complete watermark segment.
\textbf{Right:} Robustness comparison with the state-of-the-art (SOTA) watermarking tool.
Our framework demonstrates comparable watermarking capacity and imperceptibility to the leading watermarking tool while showcasing superior robustness across ten attack scenarios.
	} 
	\label{fig:overview}
\end{figure}

% \begin{figure}[th]
% 	\centering
% 	\begin{subfigure}{0.5\textwidth}
% 		\includegraphics[width=\textwidth]{figures/0_overview_v7}
% 		% \caption{Flip 1 Bit}
% 		% \label{}
% 	\end{subfigure}
%     \hspace{0.02\textwidth}
% 	\begin{subfigure}{0.4\textwidth}
% 		\includegraphics[width=\textwidth]{figures/13_radar_v0}
% 		% \caption{}
% 		% \label{fig:second}
% 	\end{subfigure}
% 	\caption{}
%  \label{fig:overview}
% \end{figure}

\section{Introduction}
\label{sec:intro}

With the growing accessibility and sophistication of voice cloning techniques~\cite{valle,valle_x,jiang2023megatts,huang2023makeavoice}, concerns about the potential misuse of synthesized speech are on the rise. Audio watermarking has emerged as a promising approach for mitigating the risks associated with voice cloning, playing an increasingly important role in safeguarding the integrity of audio recordings and ensuring that synthesized speech is used ethically and responsibly in a variety of contexts. By encoding a unique digital signature in the audio signal, audio watermarking can help verify the authenticity of a voice recording and detect any attempts to tamper with it. This technology has practical applications in 
copyright protection, broadcast monitoring, and authentication, making it a valuable tool for ensuring the accuracy and reliability of audio recordings.

Over the long term, audio watermarking has been dominated by traditional methods, such as LSB (Least Significant Bit)~\cite{LSB2004}, echo hiding~\cite{echo_hiding96}, spread spectrum~\cite{Spread_spectrum1996,Spread_spectrum1997}, patchwork~\cite{patchwork2003}, and QIM (Quantization Index Modulation)~\cite{QIM2001}.
These methods rely heavily on expert knowledge and empirical rules, 
which are challenging to implement, tending to offer a low encoding capacity while tolerating limited attacks.
In recent years, the application of deep neural networks (DNNs) in audio watermarking has shown promise~\cite{DNN2022DSP,DeAR2023AAAI}.
These works typically adopt an Encoder - Attack Simulator - Decoder  architecture, 
which can automatically learn the robustness to predefined attacks, 
significantly reducing the complexity of designing encoding strategies.
However, DNN-based audio watermarking is still in its early stage, facing problems of low capacity~\cite{DNN2022DSP} and suboptimal imperceptibility~\cite{DeAR2023AAAI}.
Furthermore, prevailing methods~\cite{DNN2022DSP,DeAR2023AAAI} solely achieved decoding on a provided watermark segment.
While in real-world scenarios, the premise of decoding is to precisely locate the watermark's position.
Thus external positioning techniques (e.g. the traditional synchronization code~\cite{sync2002}) should be involved, which complicate the implementation and may compromise the system's realiability~\cite{sync2002}.
% Thus external positioning techniques (e.g. the traditional synchronization code~\cite{sync2002}) should be involved for them to be applied to real scenes.
% However, the reliability of the localization method may become the weak link of the system~\cite{sync2002}.
% However, which may potentially compromise the realiability of the overall system~\cite{sync2002}.
% vulnerabilities that compromise the overall system~\cite{sync2002}.
% The neglect of the watermark locating problem makes them can not be directly applied to real scenes.
% Furthermore, existing DNN-based methodologies
% Furthermore, these methods need the precise location of the watermark segment to perform decoding.
% Addressing scenarios involving clipping demands the incorporation of external positioning techniques (such as traditional synchronization code~\cite{sync2002}), 
% potentially introducing vulnerabilities that compromise the overall system~\cite{sync2002}.

This paper proposes an audio watermarking framework named WavMark. As illustrated in Figure~\ref{fig:overview},
it takes 1-second audio as the host and encodes 32 bits of information.
The watermark is imperceptible to human,
% exhibits resistance to removal
and demonstrates robustness against ten common attack scenarios. A summary of our advancements over prior approaches is presented in Table~\ref{tab:main_diff}.

\begin{table}[h]
	\centering
	\caption{
		A comparison of WavMark with current DNN-based audio watermarking techniques.  Bps: bit per second.
	} 
	\label{tab:main_diff}
\begin{tabular}{ccc} 
	\toprule
	& Current DNN-Based Methods   & WavMark                   \\ 
	\hline
	Imperceptibility    & medium (SNR < 30 dB)    & high (SNR > 36 dB)        \\
	Capacity            &  < 10 bps               & 32 bps                 \\
	% Clipping Resistant~ & $ \times$                     & \checkmark                      \\
  Watermark Locating & $ \times$                     & \checkmark                      \\
	Training Data        & single type, < 1k hours & multi-types, 5k hours  \\
	\bottomrule
\end{tabular}

\end{table}
Firstly, we pioneer the application of invertible neural networks~\cite{dinh2014nice,realNVP2016} into audio watermarking.
In this framework, the encoding and decoding are regarded as reciprocal processes and share the same parameters.
This property leads to superior watermarking quality, 
making our model achieve three times the encoding capacity while preserving better imperceptibility.
Secondly, our model can automatically locate the watermark position without the hassle of external locating techniques.
It not only reduces the complexity of the system implementation but also enhances reliability.
Thirdly, unlike previous works that use single-type and hundreds of hours of data for training,
our model is trained on a diverse dataset of 5k hours, encompassing speech, music, and event sounds.
This dataset enriches our model's adaptability and helps it apply to unknown domains, such as the outputs of audio generation models~\cite{valle,musicgen,Spear_TTS}.
% Secondly, our model naturally copes with the clipping scenarios without the hassle of introducing external positioning solutions.
% Capitalizing on the solid encoding capacity of the invertible neural network, 
% our model supports decoding using in-accurate watermark position.
% This characteristic allows us to determine the watermark location through brute force detection.
% It not only reduces the complexity of the watermarking system but also enhances positioning reliability.
% Secondly, 我们的模型解决了以往工作中被忽视的水印定位问题，这使得它存在更多应用场景（例如clip）
% Secondly, our model solves the watermark location problem that has been neglected in previous works.
% which allows our model to decode any given piece of audio.
% our model realized decoding using in-accurate watermark position.
% This characteristic allows us to determine the watermark location through Brute Force Detection (BFD).

Extensive experiments demonstrate the effectiveness of our framework.
In comparison to previous DNN-based solutions,
our model achieves higher imperceptibility (↑6 dB in SNR) and double levels of robustness.
During the watermark locating test, our model shows better stability.
The average decoding error caused by imprecise localization is a mere 0.54\% BER, which is 1/19 of the traditional locating approach.
Compared to Audiowmark~\footnote{https://github.com/swesterfeld/audiowmark}, 
the SOTA open-sourced watermarking tool, our model demonstrates much higher robustness while preserving comparable capacity and imperceptibility. 
Remarkably, with 10 to 20 seconds of audio for encoding, 
our model achieves an average BER of 0.48\%, 
which is a twenty-eight-fold enhancement in robustness.

Our contributions can be summarized as follows:

1) In order to effectively address the misuse challenges associated with advanced audio generation models, we present a pioneering solution that employs invertible neural networks for audio watermarking.  By utilizing invertible neural networks, we ensure the watermark's resilience against manipulation while maintaining its inaudibility to human listeners. 

2) We propose a novel solution to the watermark localization problem, which offers simplicity in implementation and exhibits exceptional stability.
% in clipping scenarios.

3) Our work introduces various training and implementation strategies, including curriculum learning, weighted attack handling, and repeated encoding. 
These techniques collectively mitigate training complexities and improve implementation effectiveness. 
Coupled with the advanced framework structure, 
our model attains an impressive encoding capacity of 32 bits.
Moreover, the proposed model maintains commendable imperceptibility (SNR=36.85 and PESQ=4.21) while demonstrating remarkable resilience against comprehensive attacks.

\section{Related Work}
%Audio Watermark 与不可能三角

\subsection{Audio Watermarking}
Audio watermarking is an essential technology for copyright protection and content authentication, which can be dated back to the 1990s~\cite{boney1996digital,cox1997secure}. 
In the past nearly 30 years, audio watermarking technology has been dominated by traditional methods, such as echo hiding~\cite{echo_hiding96}, patchwork\cite{patchwork2000,patchwork2003}, spread spectrum~\cite{cox1997secure},
and quantization index modulation~\cite{QIM2001}.
These methods often rely on expert knowledge, empirical rules, and various heuristics, 
which are difficult to implement, fragile, and have limited adaptability.
In recent years, deep learning has achieved significant success in visual steganography~\cite{img_steganography2021,xu2022robust,bui2023rosteals} and watermarking~\cite{liu2019novel,ma2022towards,luo2023irwart}, surpassing traditional methods in encoding capacity, invisibility, and robustness. 
With the powerful modelling capabilities of DNNs, 
these models can automatically learn robust encoding methods against attacks, significantly reducing the complexity of designing watermarking strategies.
Recently, some works have extended the DNN framework to audio watermarking~\cite{DeAR2023AAAI,DNN2022DSP}.
However, they only focus on watermaking within a single watermark segment. 
The problem of watermark locating is neglected in these works,
making them can not be applied to real-world scenarios directly.

Watermark locating has been a longstanding challenge in the realm of audio watermarking~\cite{sync2001,sync2002,sync2006}.
In traditional audio watermarking, 
a marker known as the ``synchronization code~\cite{sync2002,sync2006}'' is usually added preceding the watermark segment to enable fast localization.
However, this method has proven vulnerable to desynchronization attacks (e.g., speed variation), which compromises the overall robustness of the watermarking system.
We believe that the traditional synchronization code is not a sufficient solution for DNN-based audio watermarking. 
Although DNN models can automatically obtain encoding strategies, the design of synchronization code still leans heavily on expert insight,  which divergences from achieving end-to-end audio watermarking systems. 

This paper introduces the Brute Force Detection (BFD) method as a novel solution to the watermark locating problem. 
The essence involves constructing the encoded message by combining pattern bits and payloads. 
During decoding, the model slides along the audio and continuously attempts decoding. 
The pattern bits are used as the criterion for checking the validity of the decoded outputs.
To enhance detection efficiency, we introduce a shift module into our framework.
In training, this module introduces random temporal shifts to the watermarked audio,
enforcing the model to decode using inaccurate watermark position.
As a result, our model becomes capable of decoding the watermark as long as the decoding position falls within 10\% Encoding Unit Length (EUL) distance from the actual watermark position.
If we adopt a 5\% EUL sliding step for decoding, 
merely 20 detections are required within an EUL distance, 
making the BFD method feasible for practical implementation.
% In practice, we set the maximum shift length as 10\% EUL in training, 
% and use a 5\% EUL detection step when decoding. 
% Consequently, merely 20 detections are required within an EUL distance, 
% making the BFD method feasible and efficient for practical implementation.

\subsection{Invertible Neural Network}
The concept of the invertible neural network (INN) was first introduced by Dinh~\cite{dinh2014nice} in 2014, 
which has been improved by subsequent studies like Real NVP~\cite{realNVP2016} and Glow~\cite{kingma2018glow}.
INN consists of two processes: forward and backward. 
The forward process maps complex data distributions to simple latent distributions through invertible transformations, while the backward process generates data distributions from simple latent distributions.
Due to the ability of learning direct invertible mapping, 
INN has gained extensive attention in various fields,
including image-to-image translation~\cite{van2019reversible}, 
super-resolution reconstruction~\cite{zhang2022enhancing},
visual steganography~\cite{HiNet,mou2023large},
and text-to-speech~\cite{waveglow2019,glowtts,vits}
% In the field of TTS (Text To Speech), INN has also been widely explored.
% Basing on the insights of Glow~\cite{kingma2018glow}, 
% WaveGlow~\cite{waveglow2019} ralized generaling high quality speech from melspectrograms.
% Glow-TTS~\cite{glowtts} levaraged the ﬂow-based generative model for parallel TTS.
% And VITS~\cite{vits} adopts variational inference augmented with normalizing ﬂows to realize end-to-end TTS.
% 然而，据我们所知可逆网络尚未被语音水印所应用。
% todo:可以继续调整
We believe that watermark encoding and decoding are naturally reciprocal processes.
Thus the ideal decoding should be obtained by the inverse operation of the encoding.
However, previous works~\cite{DNN2022DSP,DeAR2023AAAI} typically implemented encoder and decoder with separate networks, 
which have independent structures and parameters and cannot achieve good invertibility.
This architectural flaw makes them difficult to achieve good watermarking quality.
To the best of our knowledge, 
our work is the first application of INN in audio watermarking.

% However, we believe that invertible networks are particularly well-suited for watermarking tasks, as the watermark decoding process is precisely the inverse of the encoding process.
%And we think that invertible networks are particularly suitable for watermarking.
%Because the watermark decoding process is exactly the inverse

\section{Methods}

\begin{figure*}[h]
	\includegraphics[width=14cm]{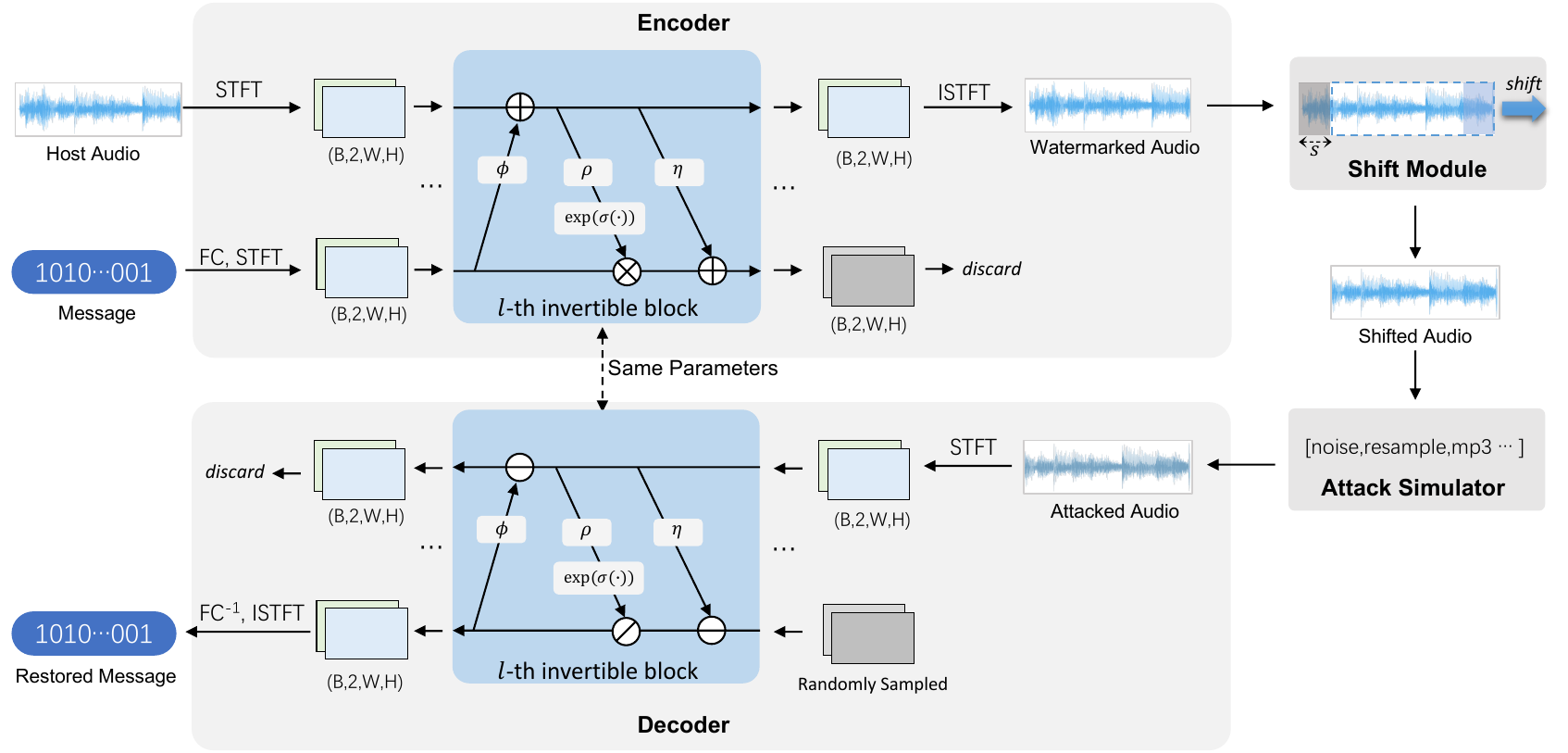}
	\centering
	\caption{
The overview of our training framework. 
The encoder combines the host audio and message vector to generate the watermarked audio. 
A shift module then randomly shifts the decoding window by a small distance. 
Random attacks are subsequently applied to the shifted audio to corrupt the watermark.
Finally, the decoder recovers the message from the attacked audio.
} 
\label{fig:model_structure}
\end{figure*}

\subsection{Overall Architecture}
The structure of our framework is depicted in Figure~\ref{fig:model_structure}, 
primarily comprising the following components: 
the invertible encoder / decoder, shift module, and attack simulator.
These modules are trained end-to-end, allowing for seamless integration and optimization. 
We provide a detailed explanation of each component in the subsequent sections.

\subsection{Invertible Encoder / Decoder}
\subsubsection{Audio Representation}
Our network takes single-channel audio with a sampling rate of 16 kHz as the host, 
where the Encoding Unit Length (EUL) is set to 1 second. 
Consequently, the original input is represented as a one-dimensional waveform vector of length 16,000,  denoted as  $\mathbf{x}_{wave} \in \mathbb{R}^L$.
We further transform it into spectrogram using the short-time Fourier transform (STFT):
\begin{equation}
	\mathbf{x}_{spec}=\Gamma _{STFT}(\mathbf{x}_{wave})
\end{equation}
This step allows us to add the watermark in the frequency domain, 
which has been shown to offer higher robustness~\cite{DNN2022DSP}.

For an input with a batch size of $B$, this process will generate a feature map $\mathbf{x}_{spec} \in \mathbb{R}^{B\times C \times W \times H } $,
where the channel $C$ is equal to 2, representing  the frequency and phase values. 
The $W$ and $H$ represent the temporal and frequency dimensions, respectively.

\subsubsection{Message Representation}
The watermark information is represented by a random binary vector of length $K$, denoted as $	\mathbf{m}_{vec} \in \{0,1\}^K$. 
We use a linear layer to expand it into a vector of the same size as the waveform input, 
and then apply the same STFT process to obtain a feature map with the same size as the  $\mathbf{x}_{spec}$:
\begin{equation}
	\mathbf{m}_{spec}=\Gamma _{STFT}(\Gamma _{FC}(\mathbf{m}_{vec})),
\end{equation}
where the $\Gamma _{FC} \in \mathbb{R}^{L \times K}$ represents a linear layer used for dimension transformation.

\subsubsection{Invertible Neural Networks}
Our invertible network is constructed by stacking $n$ layers of  invertible blocks. 
The input and output dimensions of each block maintains the same, 
and the same  parameters is used for encoding and decoding processes~\cite{realNVP2016,kingma2018glow}.
Figure~\ref{fig:model_structure} illustrates the schematic diagram of the $l$-th invertible block. 
During  encoding, 
with $\mathbf{x}^l$ and $\mathbf{m}^l$ representing the inputs of the $l$-th block, 
the outputs of the network, $\mathbf{x}^{l+1}$ and $\mathbf{m}^{l+1}$, can be described as follows:
\begin{align}
	\mathbf{x} ^{l+1} &= \mathbf{x}^l+\phi (\mathbf{m}^l)\\
	\mathbf{m}^{l+1} &= \mathbf{m}^l\odot exp(\sigma(\rho(\mathbf{x} ^{l+1}))) + \eta (\mathbf{x} ^{l+1}),
\end{align}
where the $\sigma$ denotes the sigmoid activation function, and $\odot$ represents element-wise multiplication. The functions $\phi(\cdot)$, $\eta(\cdot)$, and $\rho(\cdot)$ can be any arbitrary functions, and in this case, we utilize a dense block~\cite{DenseBlock}.

For the output of the last invertible block, 
we discard the output $\mathbf{m}^n_{spec}$ from the message branch, 
and solely utilize the output $\mathbf{x}^n_{spec}$ from the audio branch. 
Subsequently, we  perform inverse short-time Fourier transform (ISTFT) on $\mathbf{x}^n_{spec}$ to reconstruct the watermarked audio waveform:
\begin{equation}
	\mathbf{x'}_{wave}=\Gamma _{ISTFT}(\mathbf{x^{n}}_{spec})
\end{equation}

During the decoding process, 
as we only have the watermarked audio, 
we sample a variable $\mathbf{z} \in \mathbb{R}^{B\times C \times W \times H } $
from normal distribution as the input for message branch. 
The computation process for the $(l+1)$-th block in the reverse direction can be represented as follows:
\begin{align}
	\mathbf{m}^{l} &=  (\mathbf{m}^{l+1}-\eta (\mathbf{x} ^{l+1}))  \odot exp(-\sigma(\rho(\mathbf{x} ^{l+1}))) \\
	\mathbf{x}^{l} &= \mathbf{x}^{l+1}-\phi (\mathbf{m}^l).
\end{align}

Denote the spectrogram output from the message branch  as $\mathbf{\hat{m}}_{spec}$. 
After performing the ISTFT transformation, 
we obtain the frequency domain output. 
Then, by passing it through a linear layer, we can recover the message information:
\begin{equation}
	\mathbf{\hat{m}}_{vec}=\Gamma _{FC^{-1}}(\Gamma _{ISTFT}(\mathbf{\hat{m}}_{spec})).
\end{equation}

\subsection{Shift  Module}
%todo: 补充
We introduce the shift module $\Gamma_{shift}(\cdot)$ to ensure successful decoding even in the presence of displacements in the decoding position. This module is added after the encoder and shifts the decoding window along the temporal direction by a random time length $s$ (where $0 \leq s < \text{EUL}$).
We have observed that a larger shift length can weaken the encoding performance. Therefore, we restrict the maximum value of $s$ to 10\% of the EUL.
In practice, the shift operation is achieved through truncation and concatenation. We truncate the trailing $1-s$ time length from the watermarked audio and then concatenate it with the subsequent $s$ time length unwatermarked audio segment.

\subsection{Attack  Simulator}
\label{sec:attack_simulator}

%The attack  simulator $\Gamma_{attack}(\cdot)$ is employed to enhance the model's robustness against various attacks. 
%Our network design incorporates ten common attack types, including:
%Previous approaches employed complex heuristic  to achieve such balance. 

%Our primary objective is to develop a robust watermark that can withstand various common types of attacks. To achieve this, we have implemented emulation layers to counteract nine specific types of attacks, namely:

%Our goal is to make the watermark resistant to common types of attacks, 
%thus we have implemented emulation layers against the following nine attacks:\\

Our framework considers the following ten common attack types:
\begin{enumerate}
	\item Random Noise (RN): Adding a uniformly distributed noise signal to the audio, maintaining an average Signal-to-Noise Ratio (SNR) of 34.5 dB.
	\item Sample Suppression (SS): Randomly setting 0.1\% of the sample points to zero.
	\item Low-pass Filter (LP): Using a 5 kHz cutoff frequency to remove the high-frequency components in the audio.
	\item Median Filter (MF): Applying a filter kernel size of 3 to smooth the signal.
	\item Re-Sampling (RS): Converting the sampling rate to either twice or half of the original, followed by re-conversion to the original frequency.
	\item Amplitude Scaling (AS): Reducing the audio amplitude to 90\% of the original.
	\item Lossy Compression (LC): Converting the audio to the MP3 format at 64 kbps and then converting it back.
	\item Quantization (QTZ): Quantizing the sample points to $2^9$ levels.
	\item Echo Addition (EA): Attenuating the audio volume by a factor of 0.3, delaying it by 100ms, and then overlaying it with the original.
	\item Time Stretch (TS): Increasing or decreasing the speed by 1.1 or 0.9 times, respectively.
\end{enumerate}

One challenge with multiple attacks is that their learning difficulty varies. 
To address this issue, we have adopted a simple strategy of \textbf{weighted attack} to balance the learning process. 
During training, we apply only one attack type at a time. 
Initially, equal sampling weights are assigned to each attack. 
After evaluating the model on the validation set, 
we update weights with the BER values corresponding to each attack. 
As a result, attacks that are harder to learn (with higher BER) are assigned higher sampling weights, allowing the model to learn robustness against various attacks adaptively.
Furthermore, 
the sampling is performed item-wise,
enabling different attack types within the same batch to enhance training stability.

\subsection{Loss Functions}
To ensure successful decoding, we utilize L2 loss to constrain the distance between the original message $\mathbf{m}_{vec}$ and the decoded message $\hat{\mathbf{m}}_{vec}$.
Let $f$ and $f^{-1}$ represent the encoder and decoder, respectively. This loss can be written as follows:
\begin{align}
	\mathcal{L}_{m} &=||\mathbf{m}_{vec}-\hat{\mathbf{m}}_{vec}||^2_2 
	=||\mathbf{m}_{vec}-f^{-1}(\Gamma_{attack}(\Gamma_{shift}(\mathbf{x}'_{wave}),\mathbf{z})||^2_2.
\end{align}

To maintain perceptual quality, 
we first introduce a L2 constraint between the original audio $\mathbf{x}_{wave}$ and the watermarked audio $\mathbf{x}'_{wave}$:
\begin{equation}
	\mathcal{L}_a=||\mathbf{x}_{wave}-\mathbf{x}'_{wave}||^2_2=||\mathbf{x}_{wave}-f(\mathbf{x}_{wave},\mathbf{m}_{vec})||^2_2.
	\label{equ:L2_perceptual}
\end{equation}

Furthermore, we incorporate a discriminator to enhance the imperceptibility of the watermark. 
This discriminator is trained to classify host audio as 0 and the watermarked audio as 1:
\begin{equation}
	\mathcal{L}_{d}=log(1-d(\mathbf{x}_{wave}))+ log(d(\mathbf{x}'_{wave})).
\end{equation}
Our encoder network is trained to deceive this discriminator by improving audio quality, so that the watermarked audio is classified as 0:
\begin{equation}
	\mathcal{L}_{g}=log(1-d(\mathbf{x}'_{wave})).
\end{equation}

To train our watermarking network, we optimize the following total loss:
\begin{equation}
	\mathcal{L}_{total}=\lambda_a \mathcal{L}_{a}+  \mathcal{L}_{m}+\lambda_g \mathcal{L}_{g},
\end{equation}
where $\lambda_*$ controls the preference between imperceptibility and robustness. 
Simultaneously, we optimize the $\mathcal{L}_{d}$  to train the discriminator.

\subsection{Curriculum Learning Strategy}
In the initial training stage,
introducing all attacks and strong perceptual constraints makes it difficult for the model to learn effective encoding strategies. 
To address this issue, we adopt a curriculum learning approach, allowing the model to progressively acquire the encoding capabilities through three distinct stages.
In the first stage, we exclude the attack simulator and impose weak constraints on perceptibility, enabling the model to focus on learning the fundamental encoding strategy.
In the second stage, we introduce the attack simulator to enhance the model's resilience against various attacks. However, due to the weak constraints on perceptibility, the outputs might exhibit noticeable noise.
Consequently, in the third stage, we enforce strong perceptual constraints to ensure that any noise generated by the watermark remains imperceptible.

%\subsection{水印检测处理}
\section{Expirement Setting}
%The distribution of languages in this subset is illustrated in the Figure.
%In the training process, these four datasets are mixed together with the same sampling ratio.
\subsection{Datasets}
Previous works typically use single-category, hundreds of hours of data for training~\cite{DeAR2023AAAI,DNN2022DSP}. 
To enhance the model's applicability across diverse scenarios, we combined different datasets, including voice, music, and event sounds, resulting in 5,000 hours of training data.

\textbf{LibriSpeech}
LibriSpeech~\cite{panayotov2015librispeech} is an English dataset derived from the LibriVox project. We utilized the entire corpus comprising approximately 1000 hours of speech data.

\textbf{Common Voice}
Common Voice~\cite{CV2019} is a large-scale multilingual speech dataset containing over 100 languages. We selected a subset covering ten languages, totalling approximately 1,700 hours of speech data.

\textbf{Audio Set}
Audio Set~\cite{Audioset2017} is a generic audio events dataset containing 632 classes, over 5,790 hours of audio. We used a subset of 1,337 hours of data in our study.

\textbf{Free Music Archive}
Free Music Archive (FMA) ~\cite{FMA2016} is a high-quality music dataset previously used in DeAR~\cite{DeAR2023AAAI}. We leveraged the "large" subset of FMA, which includes 106,574 tracks, each lasting 30 seconds, resulting in 888 hours of music data.

For each dataset, we sampled 400 samples and reserved them for validation and testing, while the remaining was used for training.
% During training and testing,  we maintained an equal proportion of data usage for each dataset.

\subsection{Evaluation Metrics}
We use Bit Error Rate (BER) to measure the decoding accuracy, which falls within the range of [0, 1]. A BER value of 0.5 corresponds to random guessing. The calculation formula for BER is as follows:
\begin{equation}
\text{BER}(\mathbf{m}_{vec},\mathbf{m}'_{vec})=\frac{\sum_{i=1}^{K}  \mathbf{m}_{vec}(i)!=\mathbf{m}'_{vec}(i)}{K}.
\end{equation}

Following previous works~\cite{DeAR2023AAAI,DNN2022DSP}, 
we employed Signal-to-Noise Ratio (SNR) and Perceptual Evaluation of Speech Quality (PESQ)~\cite{rix2001perceptual} as evaluation metrics for audio quality.
A higher SNR value indicates better imperceptibility after watermarking:
\begin{equation}
	\text{SNR}(\mathbf{x}_{wave},\mathbf{x}'_{wave})=10 \cdot log\frac{||\mathbf{x}_{wave}||^2 }{||\mathbf{x}_{wave}(i)-\mathbf{x}'_{wave}(i)||^2} .
\end{equation}
The PESQ is based on the perceptual characteristics of the human ear, 
making it a more human-centric evaluation metric. 
PESQ scores range from [-0.5, 4.5], with values above 4.0 considered to indicate good auditory quality~\cite{DNN2022DSP}.

%相较而言，PESQ基于人耳的感知特性来评估音频质量，更加符合人类主观感知，能够较好地反映人们对音质的真实感受。PESQ取值范围为[-0.5,4.5]，一般认为PESQ>4.0 时具有较好的听觉质量。
%https://en.wikipedia.org/wiki/Perceptual_Evaluation_of_Speech_Quality

%with $\phi(\cdot)$, $\rho(\cdot)$, and $\eta(\cdot)$ functions in each block constructed using five layers of 2D convolutions.
\subsection{Implementation Details}
When performing STFT, we use the Hamming window function with a window size of 1,000 and a hop length of 400. 
For the 16,000-length waveform input, this configuration generates feature maps of size $2 \times 501 \times 41$.
Our invertible network comprises eight invertible blocks.
The $\rho(\cdot)$, $\phi(\cdot)$, $\eta(\cdot)$ functions in each invertible block are implemented as five-layer 2D CNNs with dense connections.
The discriminator architecture is designed with four layers of 1D CNNs.

During training, we set the message length $K$ as 32, resulting in an encoding capacity of 32 bps.
This model is trained on eight V100 graphics cards with the Adam optimizer. 
The curriculum learning strategy is applied, which has three stages.
In the first stage, 
we set a learning rate of $10^{-4}$, $\lambda_a=100$, $\lambda_g=10^{-4}$, and train for 3500 steps.
In the second stage, we introduce the attack simulator and continue training for 8,000 steps.
In the third stage, we decrease the learning rate to $10^{-5}$ and increase the weight of the perception loss with $\lambda_a=10^4$ and $\lambda_g=10$. Then we train the model for 57,850 steps.

\section{Experiment Results}

In this section,
we evaluate our model on different data types to reveal our model's performance in ideal segment-based and more complex utterance-based scenarios.
% and the  watermark localization ability.

\subsection{Segment based Evaluation}
\label{sec:segment_eval_with_dnn_model}
We first conduct the segment-based evaluation, where the model is required to watermark single-segment-length audios while remaining robust against various types of attacks and maintaining imperceptibility.
\begin{table*}[th]
	\centering

 	\caption{
Comparison with existing DNN-based methods.
MEAN: the average BER value across all attack scenarios (including the `No Attack' setting).
		 } 
	\label{tab:comare_DeAR_DNN22}
 
	\resizebox{1.0\columnwidth}{!}{%
\begin{tabular}{lrrrrrrrrrrrrrrr} 
\toprule
\multicolumn{1}{c}{\multirow{2}{*}{Models}} & \multicolumn{1}{c}{\multirow{2}{*}{BPS(↑)}} & \multicolumn{1}{c}{\multirow{2}{*}{SNR(↑)}} & \multicolumn{1}{c}{\multirow{2}{*}{PESQ(↑)}} & \multicolumn{12}{l}{BER(\%)(↓)}                                                         \\
\multicolumn{1}{c}{}                        & \multicolumn{1}{c}{}                        & \multicolumn{1}{c}{}                        & \multicolumn{1}{c}{}                         & MEAN & No Attack & RN   & SS   & LP   & MF   & RS   & AS   & LC   & QTZ  & EA   & TS    \\ 
\hline
\multicolumn{16}{l}{\textit{BPS<=2}}                                                                                                                                                                                                                                              \\ 
~ ~ ~RobustDNN~\cite{DNN2022DSP}                              & 1.3                                         & 24.48                                       & 3.78                                         & 0.10 & 0.00      & 0.00 & 0.00 & 0.14 & 0.42 & 0.25 & 0.00 & 0.00 & 0.00 & 0.00 & 0.12  \\
~ ~ ~WavMark-2bps                              & 2                                           & 35.65                                       & 4.21                                         & 0.08 & 0.00      & 0.00 & 0.00 & 0.00 & 0.25 & 0.06 & 0.00 & 0.00 & 0.50 & 0.00 & 0.06  \\ \\
\multicolumn{16}{l}{\textit{BPS=9}}                                                                                                                                                                                                                                              \\
~ ~ DeAR~\cite{DeAR2023AAAI}                                    & 8.8                                         & 26.18                                       & 3.73                                         & 0.48 & 0.39      & 0.39 & 0.39 & 0.96 & 0.37 & 0.38 & 0.39 & 0.45 & 0.37 & 0.42 & 0.69  \\
~ ~ WavMark-9bps                               & 9                                           & 31.24                                       & 4.09                                         & 0.12 & 0.00      & 0.03 & 0.03 & 0.03 & 0.06 & 0.04 & 0.03 & 0.00 & 0.75 & 0.03 & 0.24  \\ \\
\multicolumn{16}{l}{\textit{BPS=32}}                                                                                                                                                                                                                                              \\
~ ~ WavMark-32bps                              & 32                                          & 38.55                                       & 4.32                                         & 2.35 & 0.65      & 2.16 & 1.32 & 1.46 & 6.41 & 1.58 & 0.71 & 1.19 & 3.40 & 0.82 & 4.65  \\
\bottomrule
\end{tabular}
}

\end{table*}

We compare our model with current DNN-based methods: RobustDNN~\cite{DNN2022DSP} and DeAR~\cite{DeAR2023AAAI}.
RobustDNN achieved an encoding capacity of 1.3 bps~\footnote{
Although the RobustDNN~\cite{DNN2022DSP} successfully encoded 512-dimensional bit vectors into 2s of audio, these vectors only have six types~\cite{DNN2023}, equivalent to $(\log_{2}{6}) / 2= 1.3$ bps capacity.}, while DeAR achieved a capacity of 8.8 bps.
Since watermarking techniques face trade-offs between capacity, imperceptibility, and robustness. 
To facilitate comparisons, we additionally trained two models of 2 bps and 9 bps.
Specifically, 
we leverage the parameters of our 32 bps model as the foundation, 
reduce the message vector length $K$ to 2 and 9, respectively, 
and then fine-tune them with relaxed imperceptibility constraints.
In this evaluation, we used 400 test samples from our 5k hours dataset. 
Since the datasets and attack types used in the compared works vary.
To maintain consistency, we fine-tuned them using our dataset and attacks based on the officially released models.
The results are presented in Table~\ref{tab:comare_DeAR_DNN22}.

Within each capacity group, 
a higher imperceptibility (indicated by ↑SNR and ↑PESQ) and lower decoding error (↓BER) represent superior watermarking quality.
Compared with RobustDNN~\cite{DNN2022DSP}, our WavMark-2bps model achieves better capacity, imperceptibility and robustness simultaneously.
It has over 10 dB improvement in SNR metric while achieving lower BER in Median Filter (MF), Re-Sampling (RS), and Time Stretching (TS) metrics.
Compared to DeAR~\cite{DeAR2023AAAI}, our WavMark-9bps model also shows advantages. 
The watermarked audio not only has higher imperceptibility (↑5.06 in SNR, ↑0.36 in PESQ)
but also achieves greater robustness nearly under all attacks (except for the Quantization (QTZ) metric). 
This validates the superiority of our invertible structure in achieving higher watermarking quality.

Comparing our different capacity models,
we can see the compromise in robustness to achieve higher capacity and imperceptibility.
To achieve 38.55 dB SNR and 32 bps capacity, our 32 bps model reaches 2.35\% BER under the MEAN metric.
However, this compromise is usually acceptable in real applications, where multiple watermark segments will be added to the host to form redundancy, and error correction strategies can be further leveraged to improve robustness.

\subsection{Utterance based Evaluation}
\label{sec:real_senario_test}

\begin{figure}[h!]
	\includegraphics[width=12cm]{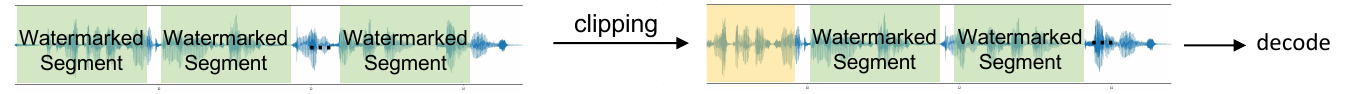}
	\centering
	\caption{
Diagram of the utterance-based evaluation.
The watermark tool adds multiple segments to the host.
Then we destroy the first watermark segment by clipping and subsequently utilize the remaining audio for decoding.
The yellow area represents the incomplete watermarked segment.
	} 
	\label{fig:task_desc_audiowm}
\end{figure}

\textbf{Setup: } 
% Audiowmark is a popular open-source watermarking toolkit developed by Stefan Westerfeld. 
Since audio in real applications varies in length,
multiple watermark segments are added to achieve full-time region protection and the  model should locate the watermark positions before decoding.
Thus we conducted an utterance-based evaluation to assess the performance of the watermark model in real-world scenarios. 
In this evaluation, 80 audios are sampled as hosts, ranging from 10 to 20 seconds each.
We experiment with two setups: clip and no-clip.
In the clip configuration, we randomly cropped between 0 and 1 second from the beginning of the watermarked audio, destroying the first watermark segment to simulate the audio cropping senario.
While in the no-clip configuration, we perform decoding directly after encoding.

As RobustDNN and DeAR cannot directly be applied to the utterance-based scenario,
we compared our model with the Audiowmark~\cite{westerfeld2020audiowmark}, the SOTA open-sourced watermarking toolkit. 
Audiowmark relies on the traditional patchwork-based watermarking~\cite{Digital2015}.
It encodes bit information by adjusting the amplitude relationships of  frequency bands and utilizes BCH codes~\cite{bose1960class} for further error correction.
We used this toolkit's default setting,
which repeatedly encodes the same 128-bit watermark into the host,
and the results of multiple segments are used for joint error correction.
Using the 16kHz audio as the host, we found a successful encoding requires an average length of 6.5 seconds.
This means that the practical encoding capacity is around 20 bps.

We used our 32 bps model for comparison with Audiowmark. 
The first 10 bits were utilized as the pattern, and the remaining 22 bits served as the payload, offering a comparable capacity as Audiowmark.
When encoding, the same watermark was repeatedly added to the host, 
with a 10\% EUL interval reserved between each segment. 
After encoding each segment, we calculate the SNR of the watermarked result.
If the SNR is higher than 38 dB, we perform \textbf{repeated encoding} (detail described in Section~\ref{sec:subsets_eval}) to realize better robustness.
During decoding, we utilize the BFD method with a step size of 5\% EUL for detection.
% We keep the results of each detection step and calculate the similarity with the pattern bits.
% The payload corresponding to the highest similarity value is used as the final result.
We calculated the similarity between the decoding result and the pattern. 
The result with the highest similarity was selected as the final  output, 
and then we calculated the BER based on the payload.

% Since our model generates decoding results for any valid input, we set the first 20 bits of the watermark to a fixed pattern while the last 12 bits represented the payload. During BFD, we set the detection step as 5\% EUL, and  we calculated the similarity between the decoding result and the pattern. 
% The result with the highest matching degree was selected as the final decoding output, and then we calculated the BER on the payload.

\begin{table}[th]
	\centering
	\caption{
 Comparision results of the utterance-based evaluation.
		% Comparision with the Audiowmark framework.
%		In this test, 80 audio samples are used  as the hosts,  each time length is 20 seconds.
	} 
	\label{tab:comare_c}
	\resizebox{1.0\columnwidth}{!}{%

\begin{tabular}{lrrcrrrrlrrrrrrr} 
	\toprule
	\multicolumn{1}{c}{\multirow{2}{*}{Methods}} & \multicolumn{1}{c}{\multirow{2}{*}{SNR(↑)}} & \multicolumn{1}{c}{\multirow{2}{*}{PESQ(↑)}} & \multicolumn{1}{l}{\multirow{2}{*}{Clip}} & \multicolumn{12}{l}{BER(\%)(↓)}                                                                \\
	\multicolumn{1}{c}{}                         & \multicolumn{1}{c}{}                     & \multicolumn{1}{c}{}                      & \multicolumn{1}{l}{}                      & MEAN  & No Attack & RN    & SS   & LP   & MF   & RS   & AS   & LC   & QTZ   & EA   & TS     \\ 
	\hline
	\multirow{2}{*}{Audiowmark~\cite{westerfeld2020audiowmark}}                  & \multirow{2}{*}{35.28}                   & \multirow{2}{*}{4.35}                     & N     & 13.60 &   3.25 &    5.93 &      3.47 &       3.25 &            7.38 &      17.20 &               3.25 &  4.10 &           6.99 &   4.51 &          50.15 \\
	&                                          &                                           & Y                                         			& 14.08 &   3.17 &    6.99 &      4.19 &       3.17 &            7.44 &      18.84 &               3.17 &  4.73 &           8.37 &   5.42 &          49.39 \\[5pt]

	\multirow{2}{*}{WavMark-32bps}                  & \multirow{2}{*}{36.85}                   & \multirow{2}{*}{4.21}                     & N     & 0.50 &   0.00 &    0.00 &      0.00 &       0.09 &            2.53 &       0.04 &               0.00 &  0.00 &           0.52 &   0.00 &           1.66 \\
	&                                          &                                           & Y                                        			& 0.48 &   0.00 &    0.35 &      0.00 &       0.00 &            1.57 &       0.22 &               0.00 &  0.00 &           0.09 &   0.00 &           1.92 \\
	\bottomrule
\end{tabular}
}

\end{table}

\textbf{Results:}
The comparison results are outlined in Table~\ref{tab:comare_c}.
We make the following observations:
Firstly, the two methods perform similarly in both clip and no-clip settings, indicating that they can effectively locate the complete watermark segment and perform decoding.
Secondly, our model demonstrates comparable imperceptibility to Audiowmark, 
with a higher SNR (↑1.57\%) and slightly lower PESQ (↓0.14\%). 
However, in terms of robustness, our model outperforms significantly. 
It surpasses Audiowmark in all attack metrics and exhibits an average BER of only 0.48\% (clip setting), a mere 1/29 of Audiowmark (14.08\%).
Thirdly, Audiowmark can not resist the Time Stretch (TS) attack. 
Because this attack destroyed its synchronization mechanism, and this toolkit failed to locate the watermark.
In contrast, our model is based on the BFD method for locating and is only slightly affected (< 2\% BER) by this attack.
The above results fully demonstrate the potential of our model in real-world applications.
% In addition, our model possesses broad scalability.
% 在这个实验中我们在每个区段上嵌入的内容是相同的，但是
% Moreover, multiple segments can be combined to achieve higher payload,
% and error correction code can be further applied to achieve higher robustness.

% Secondly, the Audiowmark can not achieve 0 BER even in the No Attack senario.
% 这显示了该方法在嵌入能力上的不足。虽然这一工具提供了参数以增强鲁棒性，但是
% 此外，我们注意到我们的模型在No Attack场景下能实现0的BER，
% 虽然在Section-1中的测试中在No Attack场景是0.65\%, 这要归功于多次嵌入与重复编码
% We take a basic implementation in this experiment, and more sophisticated measures can be introduced to enhance robustness and payload capacity. 
% For example, if we use 20-second hosts for encoding, the MEAN value in the clip case drops to a mere 0.37\%.

% Table~\ref{tab:valle_musicgan} shows the results of applying our model to the outputs of three audio generation models. 
\subsection{Evaluation on Outputs of Audio Generation Models}
\begin{table}[h]
	\centering
 	\caption{
	Utterance-based evaluation on synthetic datasets.
	% Each dataset comprises 15 audios, with duration ranging from 4 to 30 seconds.
% The encoding strategy is consistent with the settings  in Section~\ref{sec:real_senario_test}.
	} 
	\label{tab:valle_musicgan}
 
	\resizebox{1.0\columnwidth}{!}{%
\begin{tabular}{lrrrrrrrrrrrrrr} 
	\toprule
	\multirow{2}{*}{Dataset} & \multicolumn{1}{c}{\multirow{2}{*}{SNR(↑)}} & \multicolumn{1}{c}{\multirow{2}{*}{PESQ(↑)}} & \multicolumn{12}{l}{BER(\%)(↓)}                                                             \\
	& \multicolumn{1}{c}{}                     & \multicolumn{1}{c}{}                      & MEAN & No Attack & RN   & SS   & LP   & MF    & RS   & AS   & LC   & QTZ  & EA   & TS    \\ 
	\hline
%  VALL-E    & 36.62 &   3.39 &   0.56 &   0.00 &    3.18 &      0.00 &       0.00 &            0.45 &       0.00 &               0.00 &  0.00 &           3.64 &   0.00 &           0.00 \\
% Spear-TTS & 37.35 &   4.00 &   0.03 &   0.00 &    0.00 &      0.00 &       0.00 &            0.00 &       0.00 &               0.00 &  0.00 &           0.00 &   0.00 &           0.23 \\
% MusicGen  & 37.87 &   4.31 &   0.03 &   0.00 &    0.00 &      0.00 &       0.00 &            0.00 &       0.00 &               0.00 &  0.00 &           0.00 &   0.00 &           0.23 \\

 VALL-E~\cite{valle}    & 36.43 &   4.04 &   0.07 &   0.00 &    0.00 &      0.30 &       0.00 &            0.00 &       0.00 &               0.00 &  0.00 &           0.00 &   0.00 &           0.30 \\
 Spear-TTS~\cite{Spear_TTS} & 37.27 &   4.20 &   0.02 &   0.00 &    0.00 &      0.00 &       0.00 &            0.00 &       0.00 &               0.00 &  0.00 &           0.00 &   0.00 &           0.15 \\
 MusicGen~\cite{musicgen}  & 37.85 &   4.34 &   0.00 &   0.00 &    0.00 &      0.00 &       0.00 &            0.00 &       0.00 &               0.00 &  0.00 &           0.00 &   0.00 &           0.00 \\
	\bottomrule
\end{tabular}
}

\end{table}

The testing samples in the above sections are derived from our 5k hours dataset, which belong to the same domain as the training data.
In order to showcase the model's ability to handle unfamiliar domains and mitigate potential misuse of the audio generation model, we expand the evaluation at the utterance level in Table~\ref{tab:valle_musicgan}.

In this evaluation, the testing data are the outputs of audio generation models~\cite{valle,Spear_TTS,musicgen}.
Specifically, VALL-E~\cite{valle} and Spear-TTS~\cite{Spear_TTS} are models for speech synthesis, while MusicGen~\cite{musicgen} is used for music generation.
For each of them,
we collected 15 audio samples from their project's demonstration page.
The audio duration ranges from 4 to 30 seconds.
We use the same encoding configuration described in Section~\ref{sec:real_senario_test}.
Although our model was never exposed to synthetic data during the training phase, it demonstrates remarkable proficiency on synthetic datasets.
With an SNR of over 36 dB,
our model achieves 0 BER on almost all attack metrics (except for Sample Suppression (SS) and Time Stretch (TS) attacks).
These results fully demonstrate the adaptability of our model in unknown domains. Additionally, our model serves as a proactive measure against the misuse of audio generation technology.

\subsection{Watermark Locating Test}
\label{sec:locating}

\begin{figure}[h!]
	\includegraphics[width=6cm]{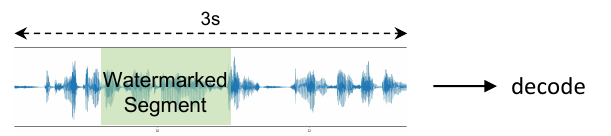}
	\centering
	\caption{
		Diagram of the watermark locating test.
	} 
	\label{fig:task_desc3s_locate}
\end{figure}

%In this section, we validate whether the proposed shift module can effectively work with BFD to address the watermark  localization problem. 

\textbf{Setup:} Since we add multiple watermarks on an utterance, how to effectively locate watermarks has been the subject of extensive research in the field over the years~\cite{sync2001,sync2002,sync2006}.
To assess the effectiveness of the proposed shift module in conjunction with BFD for addressing this problem, we conducted the watermark locating test. 
In this test, we inserted a 1-second watermark at a random position within a 3-second audio and subsequently performed decoding (Figure~\ref{fig:task_desc3s_locate}). 
Totally two hundred samples are used in this evaluation.

To compare with BFD, we introduce two settings: Oracle and SyncCode.
Oracle uses the correct positions for decoding to reflect the upper-bound performance of the localization algorithm.
As for SyncCode, we employed a traditional time-domain synchronization code~\cite{sync2002} method for localization. 
Specifically, a 12-bit Barker code sequence is added to the host right before the watermarked segment.
During decoding, the watermark location can be determined by the position with the maximum correlation between the audio and the Barker code.
% As for BFD, the first 20 bits of the watermark to a fixed pattern while the last 12 bits represented the payload. The detection step of BFD is set as 5\% EUL.

% During BFD, we set the detection step as 5\% EUL, and  we calculated the similarity between the decoding result and the pattern. The result with the highest matching degree was selected as the final decoding output, and then we calculated the BER on the payload.

\begin{table}[th]
	\centering
 	\caption{
		Comparison of different watermark localization methods.
	} 
	\label{tab:comare_oracle_bdf}
	\resizebox{1.0\columnwidth}{!}{%
		
\begin{tabular}{lrrrrrrrrrrrr} 
\toprule
\multirow{2}{*}{Locating} & \multicolumn{12}{l}{BER(\%)(↓)}                                                             \\
                          & MEAN  & No Attack & RN   & SS   & LP    & MF    & RS   & AS   & LC   & QTZ  & EA   & TS     \\ 
\hline
Oracle                    & 2.58  & 0.55      & 3.12 & 1.27 & 1.46  & 6.69  & 1.65 & 0.53 & 1.35 & 4.71 & 0.82 & 4.84   \\
SyncCode                  & 11.83 & 2.97      & 2.97 & 3.48 & 10.93 & 13.61 & 9.82 & 2.93 & 4.48 & 3.56 & 3.77 & 42.71  \\
BFD                       & 3.12  & 0.63      & 3.14 & 1.80 & 1.97  & 8.25  & 1.72 & 0.67 & 1.63 & 4.86 & 1.17 & 6.51   \\
\bottomrule
\end{tabular}

}

\end{table}

\textbf{Results:} 
The comparison results are presented in Table~\ref{tab:comare_oracle_bdf}. 
Compared with the Oracle setting,
the SyncCode locating method caused severe performance degradation.
It has over ↑9\% BER in the MEAN metric and ↑37\% BER in Time Stretch (TS).
This indicates that SyncCode becomes a weak link to the system's robustness.
Compared with SyncCode, our BFD method demonstrates superior robustness. 
When applying the BFD for locating, the BER increases are within 2\% in most attacks.
This outcome can be attributed to our framework's unique design, 
where the same model executes both decoding and localization tasks.
Consequently, these two processes have comparable resilience against diverse attacks.
This approach not only simplifies the system implementation but also enhances overall stability.

\section{Ablation Study}

\subsection{Evaluation on Different Audio Domains}
\label{sec:subsets_eval}
\begin{table}[ht]
	\centering
 \caption{
The performance of the WavMark-32bps model on each dataset (segment-based evaluation).
	} 
	\label{tab:subset_results}
	\resizebox{1.0\columnwidth}{!}{%
	
\begin{tabular}{lrrrrrrrrrrrrrr} 
	\toprule
	\multirow{2}{*}{Dataset} & \multicolumn{1}{c}{\multirow{2}{*}{SNR(↑)}} & \multicolumn{1}{c}{\multirow{2}{*}{PESQ(↑)}} & \multicolumn{12}{l}{BER(\%)(↓)}                                                             \\
	& \multicolumn{1}{c}{}                     & \multicolumn{1}{c}{}                      & MEAN & No Attack & RN   & SS   & LP   & MF    & RS   & AS   & LC   & QTZ  & EA   & TS    \\ 
	\hline
	LibriSpeech              & 39.80                                    & 4.27                                      & 0.29 & 0.00      & 0.30 & 0.06 & 0.00 & 1.29  & 0.00 & 0.00 & 0.00 & 0.72 & 0.00 & 0.68  \\
	CommonVoice              & 39.85                                    & 4.25                                      & 1.10 & 0.03      & 2.19 & 0.30 & 0.09 & 2.88  & 0.15 & 0.03 & 0.12 & 4.69 & 0.06 & 1.81  \\
	AudioSet                 & 34.94                                    & 4.33                                      & 4.18 & 1.65      & 3.76 & 2.43 & 2.85 & 10.01 & 3.00 & 1.74 & 2.43 & 4.93 & 1.95 & 8.26  \\
	FMA                      & 39.62                                    & 4.40                                      & 3.84 & 0.93      & 2.37 & 2.49 & 2.88 & 11.45 & 3.16 & 1.08 & 2.19 & 3.25 & 1.26 & 7.83  \\
	\bottomrule
\end{tabular}
}

\end{table}
Section~\ref{sec:segment_eval_with_dnn_model}'s results are the averaged value over four datasets.
In Table~\ref{tab:subset_results}, We further give the detailed performance of  WavMark-32bps model on each dataset.
It can be observed that the model achieves better performance on human voice datasets (LibriSpeech and CommonVoice),
yielding lower MEAN BER (0.29\% and 1.10\%).
While for event sounds (AudioSet) and music genres (FMA), the BER is much worse (4.18\% and 3.84\%).
The robustness decreases under all attacks, especially on the Median Filter (MF) and Time Stretch (TS) metrics.
Because the four datasets have comparable magnitudes and we perform a uniform sampling during training, this phenomenon cannot be attributed to data imbalance.
Instead, we believe that these datasets have different learning difficulties.

\begin{figure}[h]
	\includegraphics[width=14cm]{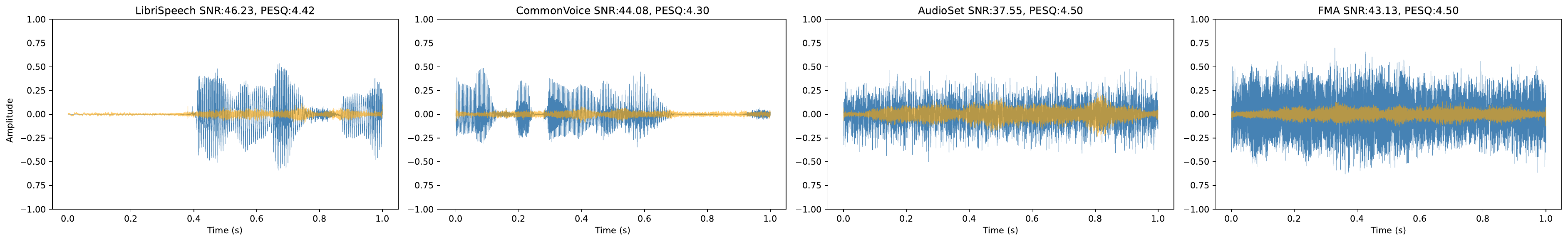}
	\centering
	\caption{
		Host audio samples. 
		The orange region represents the difference between the host audio and the watermarked audio, which is magnified by a factor of 25 for clarity ($25 \times (\mathbf{x}_{wave}-\mathbf{x}'_{wave})$).
	} 
	\label{fig:wm_diff}
\end{figure}
\begin{figure}[thb]
	\includegraphics[width=12cm]{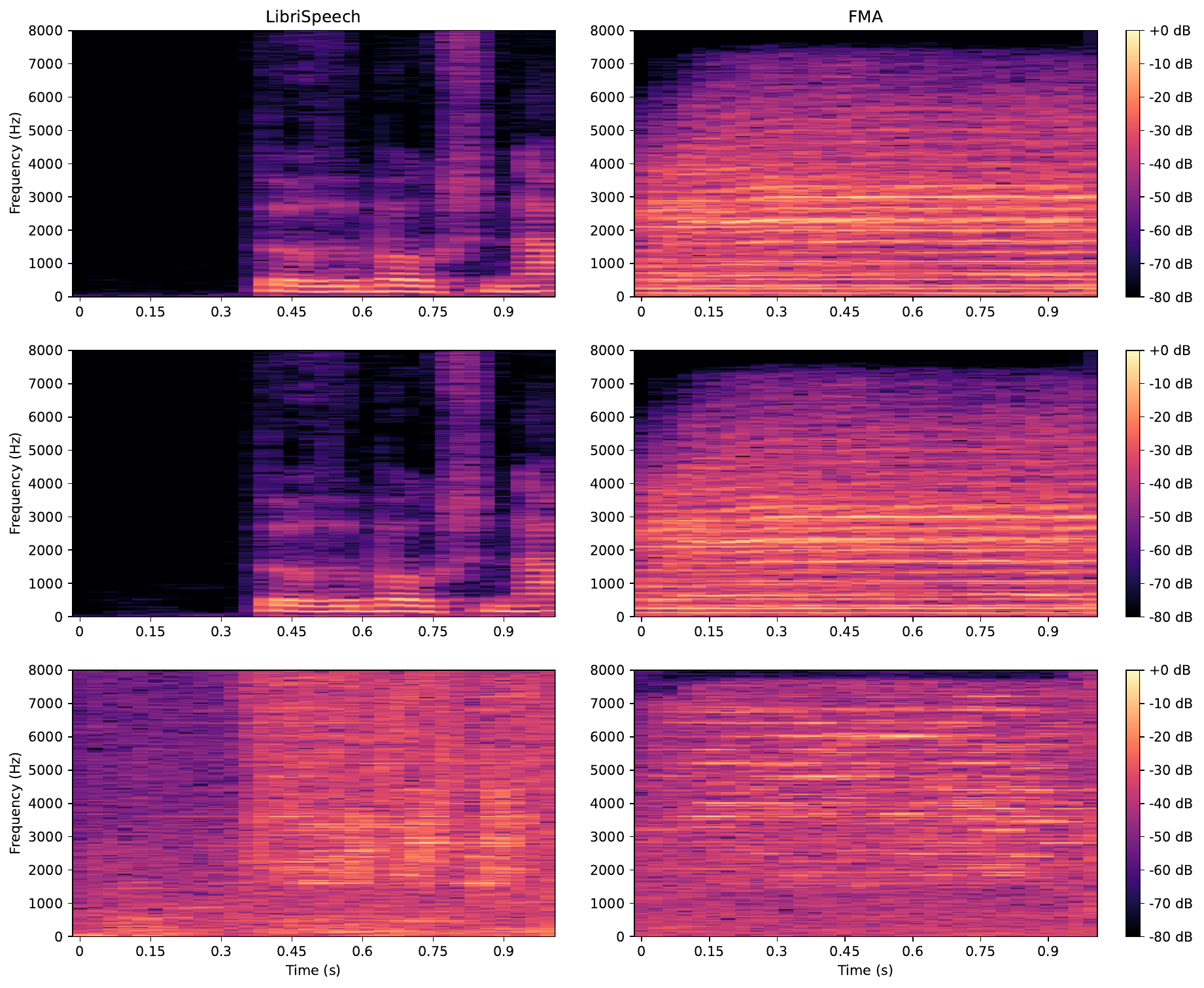}
	\centering
	\caption{
		The spectrograms of the host audio (top), watermaked audio (middle), and difference (bottom).  
	} 
	\label{fig:heatmap}
\end{figure}

We present waveform examples in Figure~\ref{fig:wm_diff} for further illustration.
The AudioSet and FMA datasets exhibit greater amplitude and fewer bass segments than LibriSpeech and CommonVoice datasets. 
An intuitive idea is that the model tends to leverage silent areas for encoding.
However, this hypothesis is not supported by Figure~\ref{fig:heatmap}, 
which shows that modifications span all time domains and the model tend to make subtle adjustments within silent areas to ensure imperceptibility.

We attribute the diminished performance on the CommonVoice and FMA datasets to their higher audio amplitudes
which need more relative modifications for successful encoding.
However, the modifications are constrained by perceptual loss (Equation~\ref{equ:L2_perceptual}), resulting in insufficient encoding and weak robustness.
To address this limitation, we integrate the concept of \textbf{repeated encoding}.
Specifically, we evaluate the SNR of the watermarked segment after the encoding.
If the encoding is insufficient (SNR > 38 dB), 
we repeat the encoding on the watermarked segment until the SNR < 38 dB. 
Through our testing, this strategy effectively improves robustness while maintaining good inaudibility (SNR > 36 dB).
Nevertheless, a potential avenue for improvement involves dynamically adjusting the level of constraint imposed by Equation~\ref{equ:L2_perceptual}.
% This adaptive adjustment could optimize the balance between inaudibility and robustness, enhancing the model's overall performance.

\subsection{Decoding Performance in Different Shifts}
% This proves that introducing the shift module enables the model to learn the shift-resistant property.
\begin{figure}[h]
	\includegraphics[width=9cm]{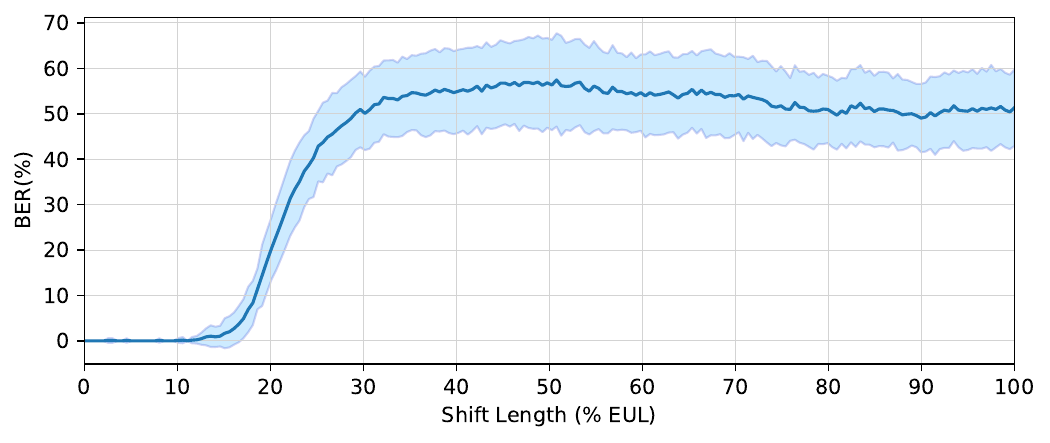}
	\centering
	\caption{
		Decoding performance in different shifts. 
		The shaded area indicates the standard deviation.
	} 
	\label{fig:ber_with_shift}
\end{figure}
Introducing the shift module enables the model to decode using inaccurate watermark location.
In Figure~\ref{fig:ber_with_shift}, we test the decoding performance in different shift lengths.
Since the max shift range is set as 10\% EUL in training, we can see the effectiveness of this setup.
When the offset < 10\% EUL, the BER value is close to 0 and has a slight standard deviation.
When the offset exceeds 10\% EUL, the model performance gradually decreases. And when the shift length reaches 35\% EUL, the performance is close to random guessing.

\subsection{Impact of the Shift Module}
\begin{table}[h]
	\centering
 	\caption{
Influence of the shift module (segment-based evaluation).
	} 
	\label{tab:shift_or_not}
 
	\resizebox{1.0\columnwidth}{!}{%

\begin{tabular}{crrrrrrrrrrrrrr} 
\toprule
\multirow{2}{*}{Shift Module} & \multirow{2}{*}{SNR(↑)} & \multirow{2}{*}{PESQ(↑)} & BER(\%)(↓) &          &       &       &       &    &    &    &    &     &    &     \\
                              &                         &                          & MEAN       & NoAttack & RN    & SS    & LP    & MF & RS & AS & LC & QTZ & EA & TS  \\ 
\hline
  N      & 40.01 &   4.33 &   1.05 &   0.30 &    0.35 &      0.57 &       0.53 &            3.73 &       0.67 &               0.30 &  0.54 &           1.45 &   0.35 &           2.11 \\
 Y      & 38.55 &   4.32 &   2.35 &   0.65 &    2.16 &      1.32 &       1.46 &            6.41 &       1.58 &               0.71 &  1.19 &           3.40 &   0.82 &           4.65 \\
\bottomrule
\end{tabular}

}

\end{table}
% todo:引入
In Table~\ref{tab:shift_or_not} we compare the impact of the shift module on the performance.
Based on the WavMark-32bps model, we remove the shift module and perform finetune to get the no-shift version. 
It can be observed that introducing the shift module affects the imperceptibility and robustness, especially on the metrics such as Random Noise (RN), Quantization (QTZ), and Time Stretch (TS). 
This implies that the shift module increases the learning difficulty of the model.
However, the it pays off in situations where the watermark localization is required.
The introduction of shift module enables us to apply the BFD method for localization, 
thus obtaining higher robustness compared to the SyncCode method (Section~\ref{sec:locating}).

% \subsection{Bit Flip}
% \begin{figure}[h]
% 	\centering
% 	\begin{subfigure}{0.45\textwidth}
% 		\includegraphics[width=\textwidth]{figures/10_figs_flip_1bit_diff}
% 		\caption{Flip 1 Bit}
% 		\label{fig:first}
% 	\end{subfigure}
% 	\begin{subfigure}{0.45\textwidth}
% 		\includegraphics[width=\textwidth]{figures/10_flip_all_diff}
% 		\caption{Flip All Bits}
% 		\label{fig:second}
% 	\end{subfigure}
	
% 	\caption{The difference caused by bit flip.}
% 	\label{fig:bit_flip}
% \end{figure}
% We apply different watermarks, namely $\mathbf{m}_{vec},\mathbf{m}_{vec}^{flip1},\mathbf{m}_{vec}^{flipK}$, to the host audio, 
% where $\mathbf{m}_{vec}^{flip1}$ and $\mathbf{m}_{vec}^{flipK}$ 
% correspond to flipping 1 bit and K bits on the message vector $\mathbf{m}_{vec}$, respectively.
% Figure~\ref{fig:bit_flip}  plots the spectrogram change of the watermarked audio caused by bit flip.
% This figure reveals that flipping a single bit induces a global change in the watermarked audio, and the change's magnitude is comparable to flipping all K bits simultaneously.

%这解释了模型鲁棒性的原因。

\begin{figure}[h!]
	\includegraphics[width=10cm]{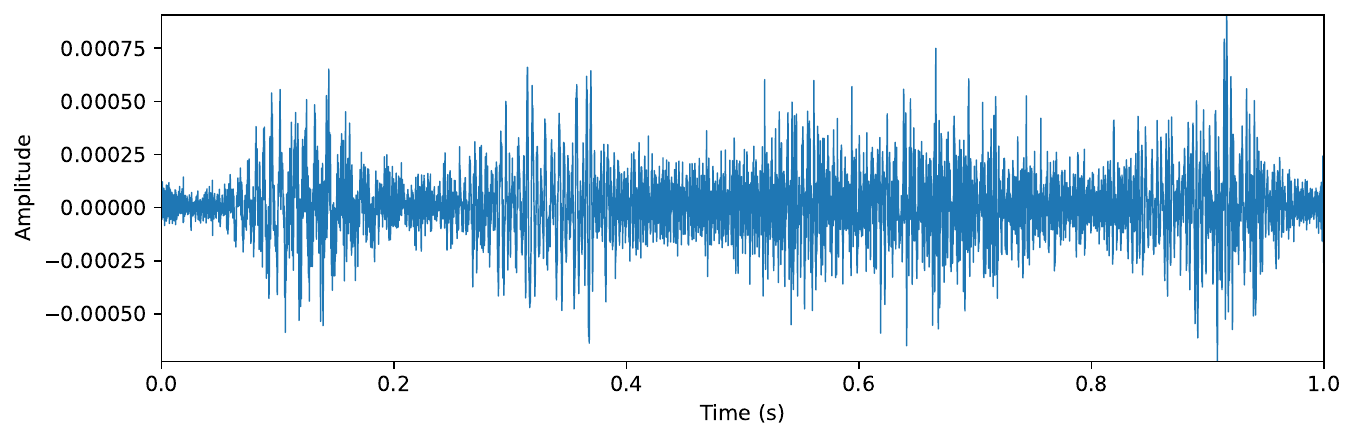}
	\centering
	\caption{
		The watermaked audio when encoding on a muted host.
	} 
	\label{fig:silent_encoding}
\end{figure}

% \subsection{Encoding on the Entirely Muted Host}

% Our model excels at achieving imperceptible encodings within non-muted audio. 
% However, there are instances where the host audio is entirely muted (e.g. the music ending).
% Therefore, we investigate the behaviour of our model when embedded into a wholly silent host segment.
% And the result of watermarked audio is depicted in Figure~\ref{fig:silent_encoding}.
% We found this watermarked audio has noticeable noise.
% This phenomenon can be attributed to our model not being explicitly trained on such data.
% In practical applications, it is advisable to omit silent segments when employing our model to ensure optimal inaudibility. 
% An efficient approach is checking the SNR after encoding and skipping the subpar quality segment (with SNR < 25dB).
% This selective approach safeguards the audio's overall imperceptibility of the watermark.

%这是合理的，因为利用掩蔽效应进行嵌入
%This signal has a peak strength of 10db, so subtle noises can be heard.

%\subsection{Repeated Embedding}
%We tested encoding a new different bit on an existing watermarked aduio.
%The results demonstrated that the new encodings effectively overwrite the old ones without  affecting the model's robustness.

\subsection{Model Size and Inference Speed}
\label{sec:speed}
For the 32 bps model, the model parameter number is 2.5 million, 
with 43\% of the parameters (1.1 million) attributed to the two linear layers in the message vector encoding and decoding process. 
With the encoding capacity increasing,  this proportion  rises accordingly. 
To address this problem, 
a viable improvement is implementing parameter-free upsampling/downsampling techniques for  feature map generation~\cite{pixinwav2023}.

On a system equipped with an AMD EPYC 7V13 CPU and an A100 GPU, 
the encoding speed is approximately 54.2 times faster than the real-time on the GPU and 7.7 times on the CPU.
And the decoding speed is comparable to the encoding. 
When utilizing BFD with a detection step of 5\% EUL, 
twenty detections will be performed within 1 EUL distance.
As a result, the BFD speed is only 0.38 times than real-time on the CPU.
However, the detection speed is typically less critical than encoding and is generally tolerable for most watermark applications.

%Although the number of parameters of the model is not high, the operation speed of the model is limited due to the large number of convolution operations involved in the invertitble blocks.

%\section{Limitations}
%While our proposed model has demonstrated superior encoding quality, 
%certain limitations offer avenues for future enhancements. 
%Firstly, 
%However, such an approach might impact the encoding quality.

\section{Conclusion and Limitations}
In this paper, we propose an invertible network-based audio watermarking framework.
It achieves 32 bps capacity, high inaudibility while maintaining robustness against ten common attacks.
In addition, we efficiently solve the localization problem overlooked in previous DNN-based studies, thus paving the way for DNN-based audio watermarking in real-world applications.

Despite the clear superiority of our proposed framework over existing DNN-based methods and established industrial solutions, there are limitations that offer valuable directions for future improvements.

\textbf{Host Audio Quality:} 
We utilized 16kHz audio as hosts. 
Extending support to higher sample rates, such as 44.1 kHz, would be crucial for accommodating a wider range of audio sources. 
However, straightforwardly increase in the host length would lead to a surge in parameter count. 
Therefore, improvements to the model's structure are necessary.

\textbf{Muted Audio:} Our model excels at achieving imperceptible encodings within non-muted audio. However, there are instances where the host audio is entirely muted (e.g. the music ending). Therefore, we investigate the behaviour of our model when encoded into a wholly silent host segment. And the result of watermarked audio is depicted in Figure~\ref{fig:silent_encoding}. We found this watermarked audio has noticeable noise. This phenomenon can be attributed to our model not being explicitly trained on such data. In practical applications, it is advisable to omit silent segments when employing our model to ensure optimal inaudibility. An efficient approach is checking the SNR after encoding and skipping the subpar quality segment (with SNR < 25 dB). This selective approach safeguards the audio's overall imperceptibility of the watermark.

\textbf{Payload Efficiency:}
Our watermark localization method relies on pattern bits to identify watermark segments. With a 32-bit capacity, if 10 bits are allocated for pattern identification, it occupies 31\% of the total capacity.
A potential improvement is to enlarge the EUL, 
which enables the encoding of additional bits and subsequently increases payload efficiency.

\textbf{Real-time Encoding:}
Currently, our model performs encodings on a fixed 1-second audio segment, requiring the presence of host audio during the encoding process. 
This constraint could pose challenges in scenarios demanding real-time watermarking.
%One possible improvement direction is to explore the host-independent watermarks~\cite{udh2020}.

%\textbf{Adaptive encoding}
%我们通过重复嵌入以适当降低不可见性为代价从而提升鲁棒性，
%一种一次性嵌入

\bibliographystyle{IEEEbib}
\bibliography{refs}

%%%%%%%%%%%%%%%%%%%%%%%%%%%%%%%%%%%%%%%%%%%%%%%%%%%%%%%%%%%%

\end{document}